\documentclass[aps,prd,twocolumn,superscriptaddress,showpacs,nofootinbib]{revtex4}
\usepackage{epsfig,epsf}
\usepackage{amsmath}
\usepackage{amsthm}
\usepackage{amsfonts}
\usepackage{amssymb}
\usepackage{dsfont}
\usepackage{multirow}
\usepackage{appendix}
\usepackage{slashed}
\usepackage[active]{srcltx}
\usepackage{psfrag}
\usepackage[xspace]{ellipsis}

\begin{document}

\title{Analysis of $P_c^+(4380)$ and $P_c^+(4450)$ as pentaquark states in the molecular picture with QCD sum rules }

\date{\today}
\author{K.~Azizi}
\affiliation{Physics Department, Do\u gu\c s University,
Ac{\i}badem-Kad{\i}k\"oy, 34722 Istanbul, Turkey}
\author{Y.~Sarac}
\affiliation{Electrical and Electronics Engineering Department,
Atilim University, 06836 Ankara, Turkey}
\author{H.~Sundu}
\affiliation{Department of Physics, Kocaeli University, 41380 Izmit, Turkey}

\begin{abstract}
To better understand the nature and internal structure of the exotic states discovered by many collaborations, more information on their electromagnetic properties and their strong and weak interactions with other hadrons is needed. The residue or current coupling constant of these states together with their mass  are the main inputs in determinations of such properties.  We perform  QCD sum rules analyses on the hidden-charm pentaquark states with spin-parities 
 $ J^{P}= \frac{3}{2}^{\pm}$  and $ J^{P}= \frac{5}{2}^{\pm}$  to calculate their  residue and mass.   In the  calculations, we adopt  a molecular picture for   $ J^{P}= \frac{3}{2}^{\pm}$ states and a mixed current in a molecular form for $ J^{P}= \frac{5}{2}^{\pm}$.  Our analyses show that the  $P_c^+(4380)$ and $P_c^+(4450)$, observed by LHCb Collaboration, can be considered as  hidden-charm pentaquark states with $ J^{P}= \frac{3}{2}^{-}$ and $ J^{P}= \frac{5}{2}^{+}$, respectively.


\end{abstract}

\pacs{12.39.Mk, 14.20.Pt, 14.20.Lq}
\maketitle

\section{Introduction}

The recent experimental progresses resulted in the observation of
exotic hadrons have placed this subject at the focus of interest.
These hadrons have an internal
structure that is more complex than those containing usual
$q\bar{q}$ or $qqq$ quark contents. The existence of these types of
hadrons is  forbidden neither in the naive quark model that
provides good description for the observed conventional hadrons nor
in quantum chromodynamics (QCD), which describes the interactions
among quarks and gluons. Starting from the observation of $X(3872)$
in 2003 by Belle collaboration~\cite{Belle} many experiments have been designed to identify
and measure the parameters of the non-conventional particles especially, the XYZ states. These experimental attempts have been accompanied by a lot of theoretical works on tetraquarks, pentaquarks, hybrids, glueballs, etc. 

The first detailed theoretical analysis on the exotic states provided by
Jaffe~\cite{Jaffe} was followed by a vast amount of theoretical
studies that investigated the properties of these particles. Among these states
are the pentaquarks for which the first claim of observation
was in 2003 through the interaction $\gamma n\rightarrow n
K^+K^-$~\cite{Nakano} suggesting a possible quark content
$uudd\bar{s}$  ($ \Theta^+ $)  with strangeness $S=+1$. Even before this claim, there
have been several works on the properties of pentaquarks (see for instance
Ref.~\cite{Gignoux,Hogaasen,Strottman,Lipkin1,Fleck,Oh,Chow,
Shmatikov,Genovese,Lipkin2,Lichtenberg}).
Later, two other experiments also found some positive signatures \cite{Barmin:2003vv,Stepanyan:2003qr}. With the
motivation provided by those results, there came another estimation on
the anti-charmed analogue of  $ \Theta^+ $ having quark content $uudd\bar{c}$ called
as $\Theta_c$. Its mass together with the mass of its b-partner $\Theta_b$ were
predicted as $2985\pm 50$ MeV and $6398\pm 50$ MeV,
respectively~\cite{Karliner:2003si}. The masses of $\Theta^+$,
$\Theta_c$ and $\Theta_b$ states were also predicted in
Ref.~\cite{Cheung:2003de}. The masses and other properties of
$\Theta^+$, $\Theta_c$ and $\Theta_b$ were then extensively examined
via various methods (see for instance 
Refs.~\cite{Matheus:2003xr,Csikor:2003ng,Oh:2003kw,Gerasyuta:2003qg,Bijker:2003pm,
Zhao:2003gs,Huang:2003bu,Nakayama:2003by,Liu:2003ab,Ko:2003xx,Huang:2004tn,Lee:2004bsa,
Kim:2004fka,Karliner:2003dt,Chiu:2004gg,Bijker:2004gr,He:2004nz,Eidemuller:2004ra,Kim:2004pu,
Melikhov:2004qh,Bicudo,Mathur:2004jr,Lee:2004xk,Nishikawa:2004tk,Kim:2005yi,Wang:2005jea,Wang:2005gv,
Eidemuller:2005jm,Wang:2005ms,Lee:2005ny,Panda:2005mj,Sarac:2005fn}
and  references therein). In the mean time, the observation
of the $\Theta_c$ was announced later by H1 collaboration at
HERA~\cite{Aktas:2004qf}. However, besides all these experiments
having positive results and related theoretical studies, some
experiments announced negative signals on the existence of those
particles~\cite{Bai:2004gk,Knopfle:2004tu,Pinkenburg:2004ux,Harris:2004kx,Karshon:2004tf,Abt:2004tz,
Aubert:2004bm,Litvintsev:2004yw,Karshon:2004kt,Link:2005ti,Aubert:2007qea}.
All these controversial results have made the
subject more intriguing from theoretical point of view since
theoretical works might provide valuable insights into the
experimental searches.

All those labor in the searches of exotic states finally resulted
in success in experimental side. With the report of observation of
$Z_c$~\cite{Ablikim:2013mio} in 2013, which might be an indication
of the existence of pentaquark states, the pentaquark captured the
attentions once more. However some experimental searches on pentaquarks
still came into view with null results such as the result of ALICE
Collaboration investigating $\phi(1869)$
pentaquark~\cite{Abelev:2014qqa} and J-PARC E19 Collaboration
searching for $\Theta^+$~\cite{Moritsu::2014qra} state. On the other hand the
theoretical studies indicated that the search on the pentaquark
containing heavy quark constituents is still
necessary~\cite{Gerasyuta:2014wka} due to the effect of such a
structure on the stability of the hadronic structures beyond the traditional hadrons~\cite{Li:2014gra}. In 2015 the
observation of two pentaquark states, $P_c^+(4380)$ and
$P_c^+(4450)$, was finally reported by LHCb Collaboration in the
$\Lambda_b^0\rightarrow J/\psi K^-p$ decays. The reported masses
were $4380\pm 8\pm 29$ MeV and $4449.8\pm 1.7\pm 2.5$ MeV with
corresponding spins $3/2$ and $5/2$ and decay widths $205\pm 18\pm
86$ MeV and $39\pm 5\pm 19$ MeV, respectively~\cite{Aaij:2015tga}.

The observation of LHCb put these particles at the focus of intense
theoretical works which aimed to explain the properties of these states. To
give an explanation on their substructure different models were
proposed. Their nature was examined using meson baryon molecular
model~\cite{Roca:2015dva,Chen:2015loa,Huang:2015uda,Meissner:2015mza,Xiao:2015fia,He:2015cea,Chen:2015moa,Chen:2016heh,Yamaguchi:2016ote,Wang:2016dzu}, diquark-triquark
model~\cite{Wang:2016dzu,Zhu:2015bba,Lebed:2015tna},
diquark-diquark-antiquark
model~\cite{Wang:2016dzu,Anisovich:2015cia,Maiani:2015vwa,Ghosh:2015ksa,Wang:2015ava,Wang:2015epa,Wang:2015ixb} and 
topological soliton model~\cite{Scoccola:2015nia}. They were also
investigated taking into account the possibility of their being a
kinematical effect or a real resonance state considering the
triangle singularity
mechanism~\cite{Guo:2015umn,Mikhasenko:2015vca,Liu:2015fea}. In \cite{oset}, however, it is concluded that with the presently claimed experimental quantum
numbers, the triangle singularity cannot be the answer for the peaks. One can
find a review on the multiquark states including pentaquarks in
Ref.~\cite{Chen:2016qju}.

All these developments make it necessary to study pentaquarks more
deeply to gain information on their nature and substructure.
The theoretical investigations on their spectroscopic and  electromagnetic properties together with their strong and weak decays may
provide valuable insights for the future experimental searches.
Moreover the comparison between new theoretical findings and
existing experimental and theoretical results may lead to a better
understanding on the nature of these particles as well as dynamics
of the strong interaction. With this motivation, in this paper, we
investigate the  residue and mass of the  hidden-charm pentaquark states with the spin-parities 
 $ J^{P}= \frac{3}{2}^{\pm}$  and $ J^{P}= \frac{5}{2}^{\pm}$. To
fulfill this aim we apply  QCD sum rule method  \cite{Shifman:1978bx,Shifman:1978by} via a choice
of interpolating current in the molecular form. Here we shall remark that the  QCD sum rule approach   in its
standard form was formulated  to reproduce the mass of the lowest hadronic state in a
given channel  with assuming that there are no other resonances close to the lowest one. We apply this method to reproduce the experimental
 data in the channels under consideration with the assumption that there are no other prominent resonances close to the lowest states with 
    $ J= \frac{3}{2}$  and $ J= \frac{5}{2}$.
In principle, there can be many interpolating currents with the same quantum numbers and  flavor contents to investigate the states  under consideration and there are no preferable interpolating currents. We  choose a molecular picture and investigate these states by considering their interpolating currents in the 
anti-charmed meson-charmed baryon  form.
For the states with $ J= \frac{5}{2}$ we consider an
admixture of $[\bar{D}\Sigma_c^*]$ and $[\bar{D}^*\Lambda_c]$  and use 
a mixed anti-charmed meson-charmed baryon molecular current. In choosing this current we consider  
the discussion given in Ref.~\cite{Chen:2016otp} which states that a choice of mixed molecular current 
provides a mass result  consistent with the experimental data.  For $ J= \frac{3}{2}$ states we also use an anti-charmed meson-charmed baryon molecular  current, namely $ \bar{D}^*\Sigma_{c} $.  As the residue is the main input in the analysis of the width, electromagnetic properties as well as the strong and weak decays of these particles, the main goal in this work is to calculate the residue of these pentaquarks with both parities considering the molecular and mixed molecular currents for $ J= \frac{3}{2}$ and $ J= \frac{5}{2}$ states, respectively.   We also calculate the mass of these states in the same pictures. Here we shall remark that in Refs. \cite{Wang:2015epa,Chen:2015moa} the authors  use the QCD sum rule method to investigate these pentaquark states, as well. In Ref. \cite{Chen:2015moa} the authors calculate only  the masses of the $ J^{P}= \frac{3}{2}^{-}$  and $ J^{P}= \frac{5}{2}^{+}$ pentaquark states with the same currents and internal quark organizations as the present work. In Ref. \cite{Wang:2015epa}, however, the author applies  diquark-diquark-antiquark type interpolating currents to calculate the mass and residue of the $ J^{P}= \frac{3}{2}^{-}$  and $ J^{P}= \frac{5}{2}^{+}$ pentaquark states.

The present work has the following outline. In Sections II and  III the
details of the mass and residue calculations for the  hidden-charm pentaquark states with $ J= \frac{3}{2}$  and $ J= \frac{5}{2}$ are presented, respectively. Section IV is set apart to the numerical analysis 
and discussion of the results. Last section is devoted to the summary and outlook.  


\section{The hidden-charm pentaquark states with $ J= \frac{3}{2}$ }


This section is devoted to present the details of the calculations on mass and residue of the  pentaquark states  with $ \frac{3}{2} $ and both the positive and negative parities.   The starting point is to consider the  following two point correlation function: 
\begin{equation}
\Pi _{1\mu \nu }(p)=i\int d^{4}xe^{ip\cdot x}\langle 0|\mathcal{T}\{J_{\mu
}^{\bar{D}^*\Sigma_{c}}(x)\bar{J}_{\nu }^{\bar{D}^*\Sigma_{c}}(0)\}|0\rangle ,  \label{eq:CorrF1Pc}
\end{equation}%
where $J_{\mu}^{\bar{D}^*\Sigma_{c}}(x)$ is the interpolating current   with $ J^{P}= \frac{3}{2}^{-} $ that couples to both the negative and positive parity particles~\cite{Chen:2015moa}:
\begin{equation}
J_{\mu}^{\bar{D}^*\Sigma_{c}}=[\bar{c}_{d}\gamma_{\mu}d_{d}][\epsilon_{abc}(u_{a}^{T}C\gamma_{\theta}u_{b})\gamma^{\theta}\gamma_{5}c_{c}].
 \label{eq:JJPc}
\end{equation}
The first step is to calculate the correlation function in terms of hadronic degrees of freedom containing the physical parameters of the states under consideration. This  requires the insertion of a complete set of the  hadronic states  into Eq.\ (\ref{eq:CorrF1Pc}) that is followed by an integration over $x$. This leads to
\begin{eqnarray}
\Pi_{1 \mu \nu }^{\mathrm{Phys}}(p)&=&\frac{\langle 0|J_{\mu }|\frac{3}{2}^{+}(p)\rangle
\langle \frac{3}{2}^{+}(p)|\bar{J}_{\nu }|0\rangle }{m_+^{2}-p^{2}} \nonumber \\&+&\frac{\langle 0|J_{\mu }|\frac{3}{2}^{-}(p)\rangle
\langle \frac{3}{2}^{-}(p)|\bar{J}_{\nu }|0\rangle }{m_-^{2}-p^{2}}+\textellipsis\ ,
\label{eq:physPc}\end{eqnarray}
where  $m_\pm$  are the masses of the positive and negative parity particles. The dots appearing in the last equation represent the contributions coming from the higher states  and continuum  resonances. The matrix elements in Eq.~(\ref{eq:physPc}) are parameterized in terms of the residues $ \lambda_+$ and  $\lambda_-$ as well as the corresponding spinors as
\begin{eqnarray}
\langle 0|J_{\mu }|\frac{3}{2}^{+}(p)\rangle &=&\lambda_+ \gamma_5 u_{\mu}(p),\nonumber\\
\langle 0|J_{\mu }|\frac{3}{2}^{-}(p)\rangle &=&\lambda_- u_{\mu}(p),
\label{eq:ResPc}
\end{eqnarray}
where  the  negative parity nature of the current under consideration has been imposed. Here we should remark that the $ J_{\mu} $  current couples not only to the spin-3/2 states, but also to the spin-1/2 states with both parities. We will choose appropriate structures to take into account only the particles with  spin-3/2. The summation
over the Rarita-Schwinger spinor is applied in the form
\begin{eqnarray}
\sum_{s}u_{\mu}(p,s)\bar{u}_{\nu}(p,s)&=&-({\slashed p}+m)\left[  g_{\mu\nu}-\frac{1}{3}\gamma_{\mu} \gamma_{\nu}\right. 
\nonumber \\ 
&-&\left. \frac{2p_{\mu}p_{\nu}}{3m^2}
+\frac{p_{\mu}\gamma_{\nu}-p_{\nu}\gamma_{\mu}}{3m}\right].
\label{eq:SumPc}
\end{eqnarray}
After the application of Borel transformation, the hadronic side gets its final form in terms of different structures, 
\begin{eqnarray}
&&\mathcal{B}_{p^{2}}\Pi_{1\mu \nu  }^{\mathrm{Phys}}(p)=-\lambda_+^{2}e^{-\frac{m_+^2}{M^2}}(-\gamma_5)({\slashed p}+m_+) 
\nonumber \\ 
&\times&
\left[  g_{\mu\nu}-\frac{1}{3}\gamma_{\mu} \gamma_{\nu}-\frac{2p_{\mu}p_{\nu}}{3m_+^2}
+\frac{p_{\mu}\gamma_{\nu}-p_{\nu}\gamma_{\mu}}{3m_+}\right]\gamma_5\nonumber\\&-&
\lambda_-^{2}e^{-\frac{m_-^2}{M^2}}({\slashed p}+m_-) 
\nonumber \\ 
&\times&
\left[  g_{\mu\nu}-\frac{1}{3}\gamma_{\mu} \gamma_{\nu}-\frac{2p_{\mu}p_{\nu}}{3m_-^2}
+\frac{p_{\mu}\gamma_{\nu}-p_{\nu}\gamma_{\mu}}{3m_-}\right] +\textellipsis\ ,
\nonumber \\  \label{eq:CorBorPc}
\end{eqnarray}
where $ M^2 $ is the Borel parameter that should be fixed later.
  To avoid the unwanted contributions coming from the spin-1/2 states, we select the $ g_{\mu\nu} $ and ${\slashed p} g_{\mu\nu} $ structures after ordering of the Dirac matrices.

To get the QCD sum rules one needs also to calculate the same correlation function  in QCD side 
in terms of quark-gluon degrees of freedom in deep  Euclidean region using the operator product expansion (OPE). This requires the contraction of the heavy and light quark fields which leads to the result
\begin{eqnarray}
&&\Pi_{1\mu \nu }^{\mathrm{QCD}}(p)=-i\int d^{4}xe^{ip\cdot x}\epsilon^{abc}\epsilon^{a^{\prime}b^{\prime}c^{\prime}}
\nonumber \\
&&\mathrm{Tr}\left[\gamma_{\mu} S_{d}^{dd^{\prime}}(x)\gamma_{\nu}S_c^{d^{\prime}d}(-x)\right] \left( \gamma{\theta}\gamma_{5}S_c^{cc^{\prime}}(x)\gamma_{5}\gamma{\beta}\right)
\nonumber \\ 
&&
\left\lbrace 
\mathrm{Tr}\left[\gamma_{\beta}\widetilde{S}_u^{aa^{\prime}}(x)\gamma_{\theta}S_u^{bb^{\prime}}(x)\right]- \mathrm{Tr}\left[\gamma_{\beta}\widetilde{S}_{u}^{ba^{\prime}}(x)\gamma_{\theta}S_u^{ab^{\prime}}(x)\right] 
\right\rbrace ,
\nonumber \\
 \label{eq:CorrF2Pc}
\end{eqnarray}%
where $ \widetilde{S}_{u(d)}(x)=CS_{u(d)}^{T}(x)C $  and the $S_{u(d)}^{ab}(x)$ and $S_{c}^{ab}(x)$ appering in Eq.~(\ref{eq:CorrF2Pc}) are the propagators of
the light $u(d)$  and  heavy $c$ quarks, respectively. The explicit expression for light quark propagator has the following form:
\begin{eqnarray}
&&S_{q}^{ab}(x)=i\delta _{ab}\frac{\slashed x}{2\pi ^{2}x^{4}}-\delta _{ab}%
\frac{m_{q}}{4\pi ^{2}x^{2}}-\delta _{ab}\frac{\langle \overline{q}q\rangle
}{12}  \notag \\
&&+i\delta _{ab}\frac{\slashed xm_{q}\langle \overline{q}q\rangle }{48}%
-\delta _{ab}\frac{x^{2}}{192}\langle \overline{q}g_{s}\sigma Gq\rangle
+i\delta _{ab}\frac{x^{2}\slashed xm_{q}}{1152}  \notag \\
&&\times \langle \overline{q}g_{s}\sigma Gq\rangle-i\frac{g_sG_{ab}^{\alpha
\beta }}{32\pi ^{2}x^{2}}\left[ \slashed x{\sigma _{\alpha \beta }+\sigma
_{\alpha \beta }}\slashed x\right]  \notag \\
&&-i\delta _{ab}\frac{x^{2}\slashed xg_{s}^{2}\langle \overline{q}q\rangle
^{2}}{7776}+\textellipsis\ ,  \label{eq:qprop}
\end{eqnarray}%
and the  heavy quark propagator is given as \cite{Reinders:1984sr}
\begin{eqnarray}
&&S_{c}^{ab}(x)=i\int \frac{d^{4}k}{(2\pi )^{4}}e^{-ikx} \Bigg \{ \frac{%
\delta _{ab}\left( {\slashed k}+m_{c}\right) }{k^{2}-m_{c}^{2}}  \notag \\
&&-\frac{g_{s}G_{ab}^{\alpha \beta }}{4}\frac{\sigma _{\alpha \beta }\left( {%
\slashed k}+m_{c}\right) +\left( {\slashed k}+m_{c}\right) \sigma _{\alpha
\beta }}{(k^{2}-m_{c}^{2})^{2}}  \notag \\
&&+\frac{g_{s}^{2}G^{2}}{12}\delta _{ab}m_{c}\frac{k^{2}+m_{c}{\slashed k}}{%
(k^{2}-m_{c}^{2})^{4}}+\textellipsis\ %
\Bigg \},  \notag \\
&& {}  \label{eq:Qprop}
\end{eqnarray}%
where we used the short-hand notations
\begin{eqnarray}
&&G_{ab}^{\alpha \beta } = G_{A}^{\alpha \beta
}t_{ab}^{A},\,\,~~G^{2}=G_{\alpha \beta }^{A}G_{\alpha \beta }^{A},
\end{eqnarray}%
in which $A=1,\,2,\,\textellipsis\ ,\,  8$ and  $a,\,b=1,2,3$ are  color indices and
$t^{A}=\lambda ^{A}/2$, with $\lambda ^{A}$ being the Gell-Mann matrices.

The  calculations in OPE side proceed by writing the correlation function in a dispersion integral form,
\begin{eqnarray}
\Pi_{1\mu\nu}^{\mathrm{QCD}}(p^{2})=\int_{(2m_{c})^{2}}^{s_0 }\frac{\rho_{\frac{3}{2}} ^{%
\mathrm{QCD}}(s)}{s-p^{2}}ds+\textellipsis\ ,
\end{eqnarray}%
where $\rho_{\frac{3}{2}} ^{\mathrm{QCD}}(s)$ is the two-point spectral density, which is found via the imaginary part of the correlation function following the standard procedures.  Here $ s_0 $ is the continuum threshold. The calculations are very lengthy. For details we refer the interested reader to e.g. Refs. \cite{Azizi:2015jya,Agaev:2016dev}. The explicit expression of the spectral density  $\rho_{\frac{3}{2}} ^{\mathrm{QCD}}(s)$, for instance for $g_{\mu\nu}$ structure, is given in the Appendix. We apply the
Borel transformation, with the aim of suppressing the contributions of the higher states and continuum, to this side  also to find the correlation function in its final form in the Borel scheme.

Now, we match  the coefficients of the structures
 $g _{\mu \nu }$ and ${\slashed p} g_{\mu\nu} $, from both the hadronic and OPE sides and apply a continuum subtraction supported by the quark-hadron duality assumption. 
 This leads to the sum rules
\begin{eqnarray}
&&m_+\lambda_+^{2}e^{-m_+^{2}/M^{2}}- m_-\lambda_-^{2}e^{-m_-^{2}/M^{2}}=%
\Pi^{1}_{1},\nonumber\\
&-&\lambda_+^{2}e^{-m_+^{2}/M^{2}}-\lambda_-^{2}e^{-m_-^{2}/M^{2}}=%
\Pi^{2}_{1},
\label{eq:srcoupling1}
\end{eqnarray}
including the masses and residues of the $\frac{3}{2}^{+}$ and $\frac{3}{2}^{-}$ states. In the last equation $\Pi^{1}_{1}$ and $\Pi^{2}_{1 }$ are the invariant functions obtained from the OPE side and correspond to the coefficients of the structures $g_{\mu\nu}$ and ${\slashed p} g_{\mu\nu}$, respectively. 

Note that Eq. (\ref{eq:srcoupling1}) contains two sum rules with four unknowns: two masses $ m_+$ and $ m_- $ as well as two residues 
$ \lambda_+ $ and $ \lambda_- $. Hence, to find these four unknowns, we need two more equations, which are found by applying the derivatives with respect to $ \frac{1}{M^{2}} $ to both sides of the above sum rules. By simultaneous solving of the four resulted equations one can find the four unknowns in terms of QCD degrees of freedom as well as the continuum threshold and Borel mass parameter.


\section{The hidden-charm pentaquark states with $ J= \frac{5}{2}$}


In this section we   follow   similar steps as  the previous section. In this case the following two point correlation function is used:
\begin{equation}
\Pi _{2\mu \nu \rho\sigma}(p)=i\int d^{4}xe^{ip\cdot x}\langle 0|\mathcal{T}\{J_{\mu\nu
}(x)\bar{J}_{\rho\sigma }(0)\}|0\rangle ,  \label{eq:CorrF1}
\end{equation}%
where $J_{\mu \nu}(x)$ is the interpolating current  having quantum numbers  $J^{P}= \frac{5}{2} ^{+}$. This current  is defined in terms of the mixed currents of $J_{\mu\nu}^{\bar{D}\Sigma_{c}^{*}}$ and $J_{\mu\nu}^{\bar{D}^{*}\Lambda_{c}}$ via the expression~\cite{Chen:2015moa}
\begin{eqnarray}
J_{\mu\nu}(x)=\mathrm{\sin\theta} \times J_{\mu\nu}^{\bar{D}\Sigma_{c}^{*}}+\mathrm{\cos\theta} \times J_{\mu\nu}^{\bar{D}^{*}\Lambda_{c}},
\label{eq:CDiq}
\end{eqnarray}
where $\theta$ is a mixing angle and
\begin{eqnarray}
J_{\mu\nu}^{\bar{D}\Sigma_{c}^{*}}&=&[\bar{c}_{d}\gamma_{\mu}\gamma_{5}d_{d}][\epsilon_{abc}(u_{a}^{T}C\gamma_{\nu}u_{b})c_{c}]+\lbrace\mu\leftrightarrow\nu\rbrace,\nonumber \\
J_{\mu\nu}^{\bar{D}^{*}\Lambda_{c}}&=&[\bar{c}_{d}\gamma_{\mu}u_{d}][\epsilon_{abc}(u_{a}^{T}C\gamma_{\nu}\gamma_{5}d_{b})c_{c}]+\lbrace\mu\leftrightarrow\nu\rbrace.\nonumber \\
 \label{eq:JJ}
\end{eqnarray}%
In Ref. \cite{Chen:2015moa} it is found that the above current with the mixing angle $ \theta=(-51\pm5)^\circ $ gives a result  consistent with the experimental mass of
$P_c(4450)$ state \footnote{Our analyses show that the results do not depend on $\theta  $ considerably. Hence,  an optimization as advised in Ref. \cite{Wang:2015uha} does not work in this case.}.

The hadronic side after integration over $x$ is obtained as
\begin{eqnarray}
\Pi _{2\mu \nu\rho\sigma }^{\mathrm{Phys}}(p)&=&\frac{\langle 0|J_{\mu \nu}|\frac{5}{2}^{+}(p)\rangle
\langle \frac{5}{2}^{+}(p)|\bar{J}_{\rho\sigma }|0\rangle }{m_+^{2}-p^{2}}\nonumber\\&+&\frac{\langle 0|J_{\mu \nu}|\frac{5}{2}^{-}(p)\rangle
\langle \frac{5}{2}^{-}(p)|\bar{J}_{\rho\sigma }|0\rangle }{m_-^{2}-p^{2}}\nonumber\\&+& \textellipsis\ ,
\label{eq:phys}\end{eqnarray}%
with $m_\pm$ being the masses of the $\frac{5}{2}$ states having positive and negative parities. The contributions of the higher and continuum states resonances to the correlation function are represented via the dots appearing in the last equation. For the matrix elements presented in Eq.~(\ref{eq:phys}) the following parameterizations in terms of the residues and spinors are used:
\begin{eqnarray}
\langle 0|J_{\mu\nu }|\frac{5}{2}^{+}(p)\rangle &=&\lambda_+u_{\mu\nu}(p),\nonumber\\\langle 0|J_{\mu\nu }|\frac{5}{2}^{-}(p)\rangle &=&\lambda_-\gamma_5u_{\mu\nu}(p).
\label{eq:Res}
\end{eqnarray}%
The current $J_{\mu \nu}(x)$  also couples to the states with spin 3/2 and 1/2 with both parities. Again, we will choose the structures that only give contributions to the spin 5/2 particles. 
With the usage of the summation~\cite{Wang:2015epa},
\begin{eqnarray}
\sum_{s}u_{\mu\nu}\bar{u}_{\rho\sigma}&=&({\slashed p}+m)\left\lbrace   \frac{\widetilde{g}_{\mu\rho}\widetilde{g}_{\nu\sigma}+\widetilde{g}_{\mu\sigma}\widetilde{g}_{\nu\rho}}{2}-\frac{\widetilde{g}_{\mu\nu}\widetilde{g}_{\rho\sigma}}{5}\right.  \nonumber \\&-&\frac{1}{10}\left(\gamma_{\mu}\gamma_{\rho}+\frac{\gamma_{\mu}p_{\rho}-\gamma_{\rho}p_{\mu}}{\sqrt{p^{2}}}-\frac{p_{\mu}p_{\rho}}{p^{2}}\right) \widetilde{g}_{\nu\sigma}
\nonumber \\&-&\frac{1}{10}\left(\gamma_{\nu}\gamma_{\rho}+\frac{\gamma_{\nu}p_{\rho}-\gamma_{\rho}p_{\nu}}{\sqrt{p^{2}}}-\frac{p_{\nu}p_{\rho}}{p^{2}} \right) \widetilde{g}_{\mu\sigma}
\nonumber \\&-&\frac{1}{10}\left(\gamma_{\mu}\gamma_{\sigma}+\frac{\gamma_{\mu}p_{\sigma}-\gamma_{\sigma}p_{\mu}}{\sqrt{p^{2}}}-\frac{p_{\mu}p_{\sigma}}{p^{2}} \right) \widetilde{g}_{\nu\rho}
\nonumber \\&-& \left. \frac{1}{10}\left(\gamma_{\nu}\gamma_{\sigma}+\frac{\gamma_{\nu}p_{\sigma}-\gamma_{\sigma}p_{\nu}}{\sqrt{p^{2}}}-\frac{p_{\nu}p_{\sigma}}{p^{2}} \right) \widetilde{g}_{\mu\rho}
\right\rbrace ,\nonumber \\
\label{eq:Sum}
\end{eqnarray}
where $  \widetilde{g}_{\mu\nu}=g_{\mu\nu}-\frac{p_{\mu}p_{\nu}}{p^{2}}$, the correlation function gets the form
\begin{eqnarray}
\Pi _{2\mu \nu \rho\sigma}^{\mathrm{Phys}}(p)&=&\frac{\lambda_+^{2}}{
( m_+^{2}-p^{2}) } (\slashed p+m_+) \frac{g_{\mu\rho}g_{\nu\sigma}+g_{\mu\sigma}g_{\nu\rho}}{2}\nonumber\\&+&\frac{\lambda_-^{2}}{
( m_-^{2}-p^{2}) } (\slashed p-m_-) \frac{g_{\mu\rho}g_{\nu\sigma}+g_{\mu\sigma}g_{\nu\rho}}{2}\nonumber\\&+&\textellipsis\  ,  \label{eq:CorM}
\end{eqnarray}%
in terms of $m_+$, $m_-$, $\lambda_+$ and $\lambda_-$. In the last result there are other Lorentz structures giving contributions to the correlation function, however  those structures mainly include contributions also coming from other pentaquark states having spin-1/2 and spin-3/2. To exclude such type of contributions,  in the remaining part of the calculations we use the presented structures to extract the mass and residue of the states under consideration. Therefore the dots in Eq.~(\ref{eq:CorM}) represents both the contributions coming from other Lorentz structures that are not written explicitly here as well as the contributions of higher states and continuum. Application of Borel transformation to Eq.\ (\ref{eq:CorM}) results in
\begin{eqnarray}
\mathcal{B}_{p^{2}}\Pi_{2\mu \nu \rho\sigma}^{\mathrm{Phys}}(p)&=&\lambda_+^{2}
e^{-\frac{m_+^2}{M^2}}
(\slashed p+m_+) 
(\frac{g_{\mu\rho}g_{\nu\sigma}+g_{\mu\sigma}g_{\nu\rho}}{2}) \nonumber\\
&+&\lambda_-^{2}
e^{-\frac{m_-^2}{M^2}}
(\slashed p-m_-) 
(\frac{g_{\mu\rho}g_{\nu\sigma}+g_{\mu\sigma}g_{\nu\rho}}{2})\nonumber\\&+&\textellipsis\ . \label{eq:CorBor}
\end{eqnarray}

In order to obtain the   QCD side of the correlation function,
we contract  the heavy and light quark fields using the Wick's theorem, which leads to 
\begin{eqnarray}
&&\Pi _{2\mu \nu \rho\sigma}^{\mathrm{QCD}}(p)=i\int d^{4}xe^{ipx}\epsilon^{abc}\epsilon^{a^{\prime}b^{\prime}c^{\prime}}\left\{ \mathrm{\sin^{2}\theta}~ S_{c}^{cc^{\prime}}(x)\right. 
\nonumber \\
&\times& 
\left.
\{ \mathrm{Tr}
\left[ \gamma_{\mu}\gamma _{5}S_{d}^{dd^{\prime}}(x)\gamma_{5}\gamma_{\rho}S_{c}^{d^{\prime}d}(-x)\right] \left( \mathrm{Tr}
\left[\gamma_{\nu}S_{u}^{ba^{\prime}}(x)\gamma_{\sigma}\right.\right.\right.\nonumber\\
&\times& \left.\left.\left. \widetilde{S}_u^{ab^{\prime}}(x)\right]-\mathrm{Tr}\left[\gamma_{\nu}S_{u}^{bb^{\prime}}(x)\gamma_{\sigma}\widetilde{S}_u^{aa^{\prime}}(x)\right]
 \right)\}
+ \mathrm{\cos^{2}\theta}\right. 
\nonumber \\
&\times& 
\left. S_{c}^{cc^{\prime}}(x)\{ 
\mathrm{Tr}
\left[ \gamma_{\mu}S_{u}^{da^{\prime}}(x)\gamma_{\sigma}\gamma_{5}\widetilde{S}_{d}^{bb^{\prime}}(x)\gamma_{5}\gamma_{\nu}S_{u}^{ad^{\prime}}(x)\gamma_{\rho}\right.\right.\nonumber\\
&\times & \left.\left.\left.S_{c}^{d^{\prime}d}(-x)\right] 
-\mathrm{Tr}\left[\gamma_{\mu}S_{u}^{dd^{\prime}}(x)\gamma_{\rho}S_c^{d^{\prime}d}(-x)\right]\mathrm{Tr}
\Big[\gamma_{\nu}\gamma_{5}\right.\right.
\nonumber\\
&\times &\left.\left. S_{d}^{bb^{\prime}}(x)\gamma_{5}\gamma_{\sigma}
\widetilde{S}_u^{aa^{\prime}}(x)\right]
 \}+\mathrm{\sin\theta}~\mathrm{\cos\theta}~ S_{c}^{cc^{\prime}}(x)\right.\nonumber\\
  &\times& 
  \left.
  \{ 
\mathrm{Tr}
\left[ \gamma_{\mu}\gamma_{5} S_{d}^{db^{\prime}}(x)\gamma_{5} \gamma_{\sigma}\widetilde{S}_{u}^{aa^{\prime}}(x)\gamma_{\nu}S_{u}^{bd^{\prime}}(x)\right.\right.\nonumber\\
&\times &\left.\left.
\gamma_{\rho}S_{c}^{d^{\prime}d}(-x)\right]-\mathrm{Tr}\left[ \gamma_{\mu}\gamma_{5} S_{d}^{db^{\prime}}(x)\gamma_{5} \gamma_{\sigma}\widetilde{S}_{u}^{ba^{\prime}}(x)\gamma_{\nu}
\right.\right.\nonumber\\
&\times &\left.\left. S_{u}^{ad^{\prime}}(x)
\gamma_{\rho}S_{c}^{d^{\prime}d}(-x)\right]+\mathrm{Tr}\left[ \gamma_{\mu} S_{u}^{db^{\prime}}(x) \gamma_{\sigma}\widetilde{S}_{u}^{aa^{\prime}}(x)
\right.\right.\nonumber\\
&\times &\left.\left.
\gamma_{\nu} \gamma_{5}S_{d}^{bd^{\prime}}(x)\gamma_{5}\gamma_{\rho}S_{c}^{d^{\prime}d}(-x)\right]-\mathrm{Tr}\left[ \gamma_{\mu} S_{u}^{da^{\prime}}(x) \gamma_{\sigma}\right.\right.\nonumber\\
&\times &\left.\left.
\widetilde{S}_{u}^{ab^{\prime}}(x)\gamma_{\nu} \gamma_{5}S_{d}^{bd^{\prime}}(x)
\gamma_{5}\gamma_{\rho}S_{c}^{d^{\prime}d}(-x)\right]\}
+\left( \mu\leftrightarrow \nu\right) 
\right.\nonumber\\ 
&+&\left.\left( \rho\leftrightarrow \sigma\right)+ \left( \mu\leftrightarrow \nu, \rho\leftrightarrow \sigma\right)\right.  
\Big\}.
 \label{eq:CorrF2}
\end{eqnarray}%
In this step we use the expressions of the heavy and light propagators and transform the calculations 
to the momentum space. By using the dispersion relation we find the imaginary part of the correlation function to extract the corresponding spectral density of $\frac{5}{2}$ state. Omitting the details of very lengthy calculations, we show the spectral density $ \rho_{\frac{5}{2}}^{\mathrm{QCD}} (s) $ defining the state under consideration, for instance for $  \frac{g_{\mu\rho}g_{\nu\sigma}+g_{\mu\sigma}g_{\nu\rho}}{2} $ structure,   in the Appendix.

By matching the coefficients of the selected structures from both sides we find the sum rules
\begin{eqnarray}
&&m_+\lambda_+^{2}e^{-m_+^{2}/M^{2}}- m_-\lambda_-^{2}e^{-m_-^{2}/M^{2}}=%
\Pi^{1}_{2},\nonumber\\
&&\lambda_+^{2}e^{-m_+^{2}/M^{2}}+\lambda_-^{2}e^{-m_-^{2}/M^{2}}=%
\Pi^{2}_{2},
\label{eq:srcoupling}
\end{eqnarray}
where $ \Pi^{1}_{2} $ and $\Pi^{2}_{2}  $ correspond to the coefficients of the structures $  \frac{g_{\mu\rho}g_{\nu\sigma}+g_{\mu\sigma}g_{\nu\rho}}{2} $ and $  \slashed p\frac{g_{\mu\rho}g_{\nu\sigma}+g_{\mu\sigma}g_{\nu\rho}}{2} $ in the OPE side, respectively. The four unknowns $ m_+$ and $ m_- $,  $ \lambda_+ $ and $ \lambda_- $ can be obtained using the above two sum rules and two extra sum rules obtained via applying  the derivatives with respect to $ \frac{1}{M^2} $ to their  both sides.


%
%


\section{Numerical results}

\label{sec:Num}
\begin{table}[tbp]
\begin{tabular}{|c|c|}
\hline\hline
Parameters & Values \\ \hline\hline
$m_{c}$ & $(1.27\pm0.03)~\mathrm{GeV}$ \\
$\langle \bar{q}q \rangle $ & $(-0.24\pm 0.01)^3$ $\mathrm{GeV}^3$  \\
$m_{0}^2 $ & $(0.8\pm0.1)$ $\mathrm{GeV}^2$ \\
$\langle \overline{q}g_s\sigma Gq\rangle$ & $m_{0}^2\langle \bar{q}q \rangle$
\\
$\langle\frac{\alpha_sG^2}{\pi}\rangle $ & $(0.012\pm0.004)$ $~\mathrm{GeV}%
^4 $\\
\hline\hline
\end{tabular}%
\caption{Some input parameters used in the calculations.}
\label{tab:Param}
\end{table}
The  QCD sum rules for the physical quantities under consideration contain some parameters such as quark, gluon and mixed
condensates  and mass of the $c$ quark.  We collect their values in Table~\ref{tab:Param}. We set the light quark masses, $ m_u $ and $ m_d $ to zero.
In addition to the above parameters, there are two  auxiliary parameters  that should be fixed before going further,  namely the continuum threshold $s_{0}$ and Borel parameter $M^{2}$. We find their working windows such  that the physical quantities under consideration be roughly independent of these parameters.  To determine the working interval of the Borel parameter one needs to consider two criteria:  convergence of the series of OPE and adequate suppression of the  higher states and continuum . Consideration of these criteria in the analysis leads to the intervals
\begin{equation}
4\ \mathrm{GeV}^{2}\leq M^{2}\leq 7\ \mathrm{GeV}^{2}.
\end{equation}%
To determine the working regions of the continuum threshold, we  impose the conditions of pole dominance and OPE  convergence.  This leads to the interval
\begin{equation}
22\,\,\mathrm{GeV}^{2}\leq s_{0}\leq 24\,\,\mathrm{GeV}^{2},
\end{equation}
for $\frac{3}{2}$ states with both  parities and
\begin{equation}
22.5\,\,\mathrm{GeV}^{2}\leq s_{0}\leq 24.5\,\,\mathrm{GeV}^{2},
\end{equation}
for $\frac{5}{2}$ states with negative and positive parities.


As examples, the variations  of the mass and residue of the hidden-charm pentaquark with $J =\frac{5}{2} $ and  positive parity with respect to the Borel parameter (continuum threshold) at different fixed values of the continuum threshold (Borel parameter) are depicted in figures  \ref{mass52Msq} - \ref{mass52s0}.   From these figures we see that the  corresponding mass and residue  demonstrate overall weak dependence on the variations of the Borel mass parameter and continuum threshold in their working intervals.  
\begin{widetext}

\begin{figure}[h!]
\begin{center}
\includegraphics[totalheight=5cm,width=7cm]{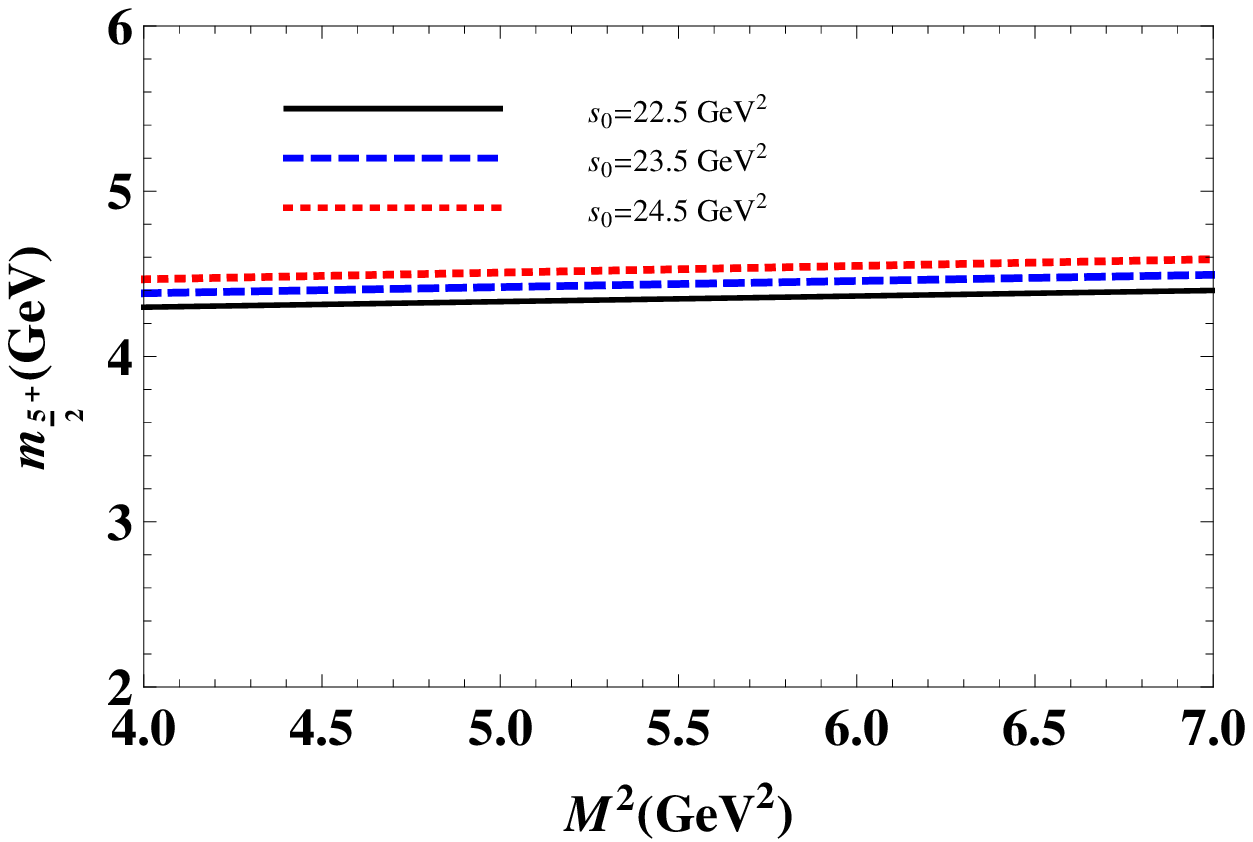}
\includegraphics[totalheight=5cm,width=7cm]{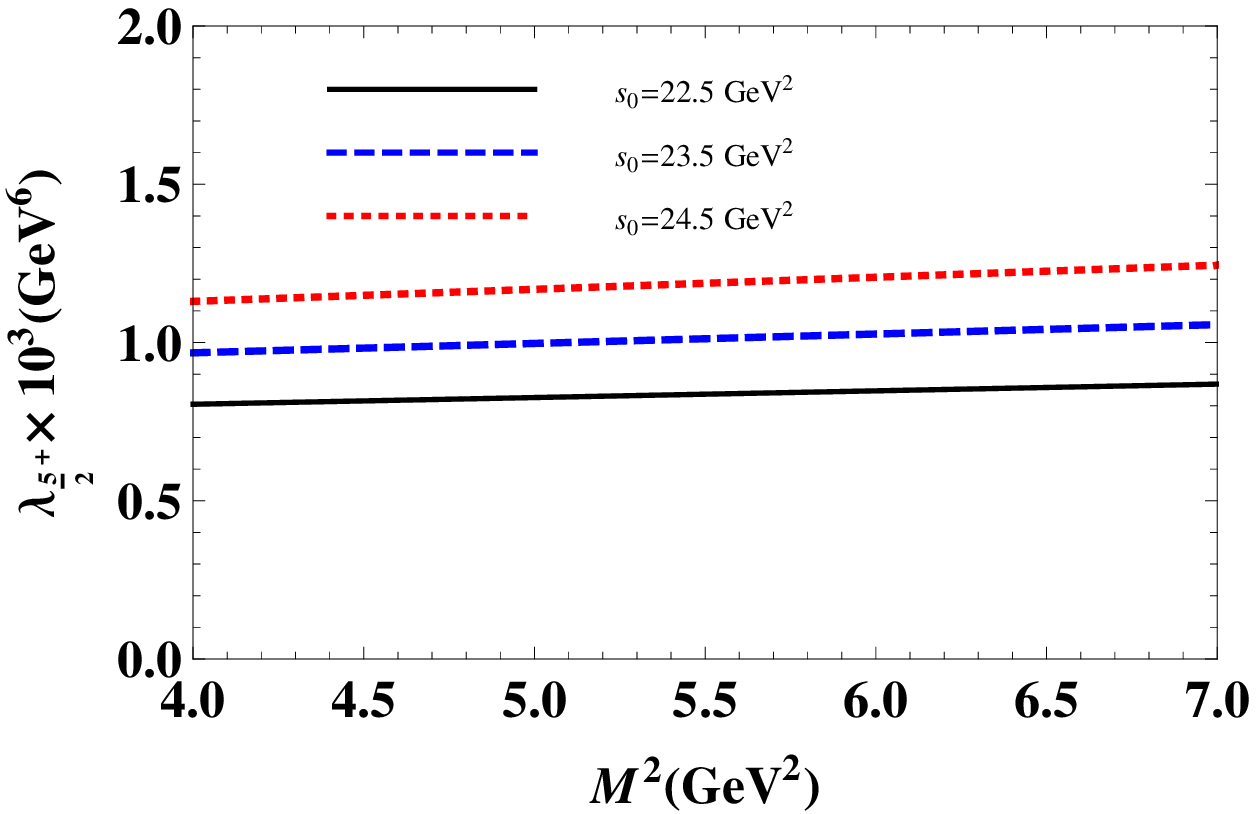}
\end{center}
\caption{\textbf{Left:} The mass  of the  pentaquark with $J^{P} =\frac{5}{2} ^{+} $  as a function of Borel
parameter $M^2$ at different fixed values of the continuum threshold. \textbf{Right:}
 The residue  of the  pentaquark with $J^{P} =\frac{5}{2} ^{+} $  as a function of Borel
parameter $M^2$ at different fixed values of the continuum threshold.  } \label{mass52Msq}
\end{figure}
\begin{figure}[h!]
\begin{center}
\includegraphics[totalheight=5cm,width=7cm]{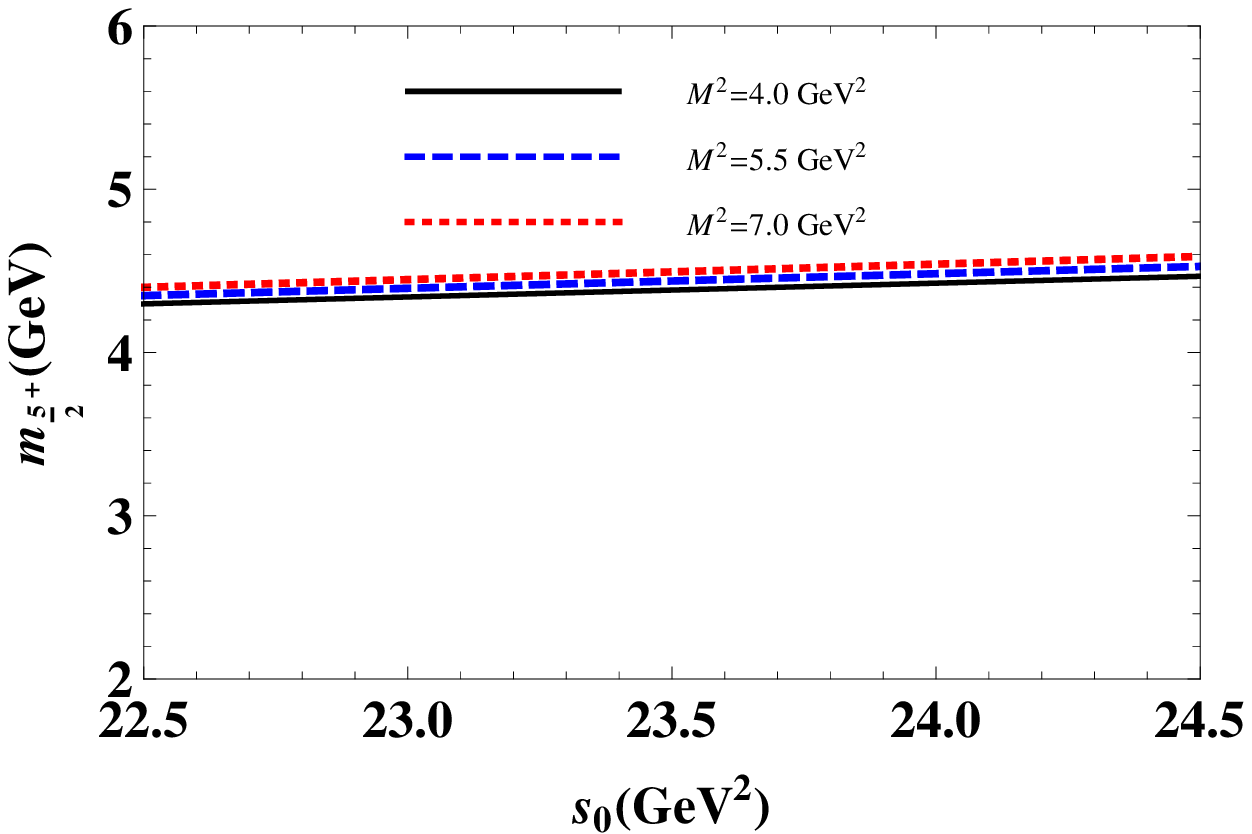}
\includegraphics[totalheight=5cm,width=7cm]{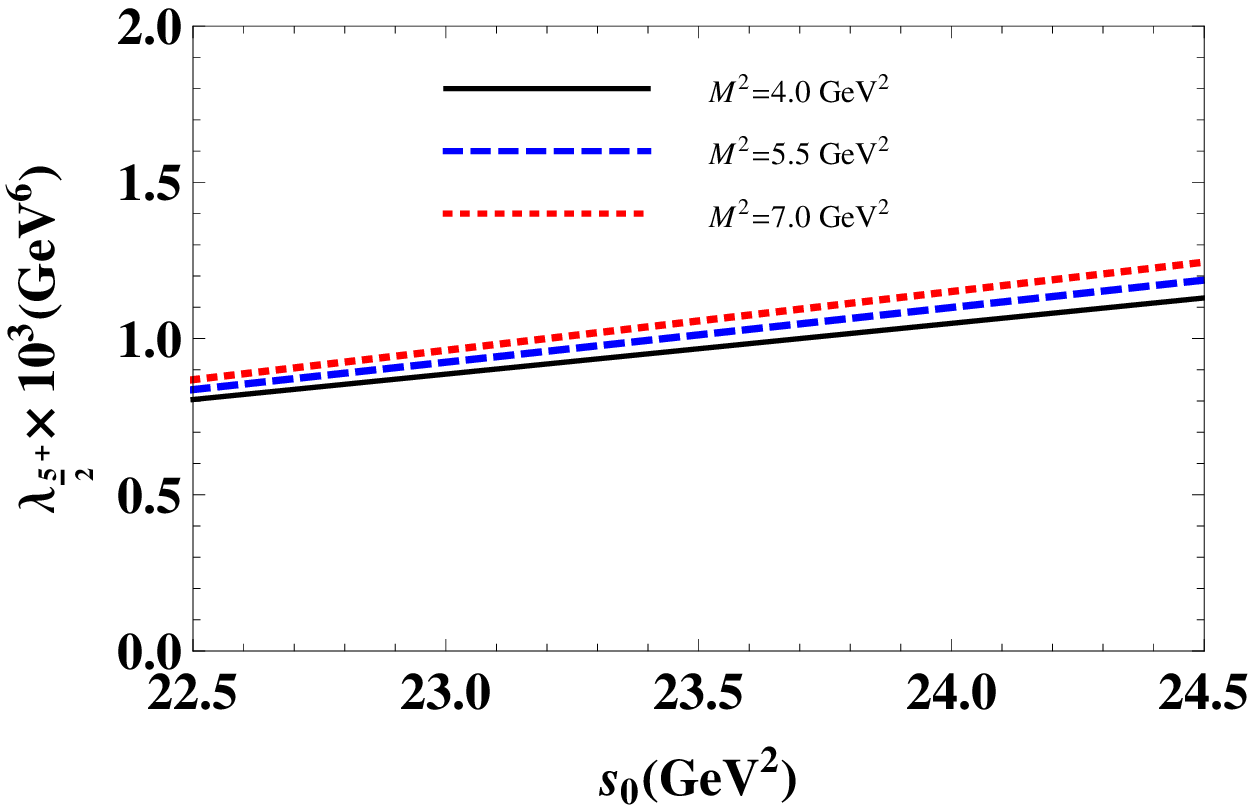}
\end{center}
\caption{\textbf{Left:} The mass  of the pentaquark with $J^{P} =\frac{5}{2} ^{+} $  as a function of  $s_0$ at different fixed values of the Borel
parameter. \textbf{Right:}
 The residue of the pentaquark with $J^{P} =\frac{5}{2}  ^{+}$  as a function of  $s_0$ at different fixed values of the Borel
parameter.} \label{mass52s0}
\end{figure}

\end{widetext}
%
%

 Having determined the suitable  intervals for the parameters $s_0$ and $M^2$ the next stage is to use them in the determination of 
 mass and residue  of the considered pentaquarks. 
 The  average values obtained from our calculations are presented in Table~\ref{tab:Values}. 
\begin{table}[tbp]
\begin{tabular}{|c|c|c|}
\hline\hline
            $ J^{P} $  &  $m~(\mathrm{GeV})$& $\lambda~(\mathrm{GeV}^{6})$ \\ \hline\hline
$\frac{3}{2}^{+}$ &  $4.24\pm 0.16$  & $ (0.59\pm 0.07)\times 10^{-3}$        \\
\hline
$\frac{3}{2}^{-}$ &  $4.30\pm 0.10$  & $ (0.94\pm 0.05)\times 10^{-3}$        \\
\hline
$\frac{5}{2}^{+}$ &  $4.44\pm 0.15$  & $ (1.01\pm 0.23)\times 10^{-3}$        \\
\hline
$\frac{5}{2}^{-}$ &  $4.20\pm 0.15$  & $ (0.51\pm 0.09)\times 10^{-3}$        \\
\hline\hline
\end{tabular}%
\caption{The results of QCD sum rules calculations for the mass and residue of the pentaquark states.}
\label{tab:Values}
\end{table}
The errors in the given results  arise due to the input parameters and also due to the uncertainties coming from the determination of the working windows of the auxiliary parameters $s_0$ and $M^2$.  Comparison of the results on the masses with the experimental data of   LHCb Collaboration, i.e.,  $m_{P_c^+(4380)}=4380\pm 8\pm 29$~MeV and $m_{P_c^+(4450)}=4449.8\pm 1.7\pm 2.5$~MeV~\cite{Aaij:2015tga} reveals that the $\frac{3}{2}^{-}$ state can be assigned to  $P_c^+(4380)$  observed by LHCb. Our prediction for the mass of $\frac{5}{2}^{+}$ state  is also consistent with the experimental data on the mass of the $P_c^+(4450)$. Our results on the masses of the $\frac{3}{2}^{-}$  and $\frac{5}{2}^{+}$  are also in a good agreement with the results of the theoretical works~\cite{Wang:2015epa,Chen:2015moa}. 
Our predictions on the residues of the  $\frac{3}{2}^{-}$ and $\frac{5}{2}^{+}$ states, within the errors,   are also comparable with the predictions of   \cite{Wang:2015epa}, which applies  diquark-diquark-antiquark type interpolating currents to calculate the mass and residue of the pentaquark states with $J^{P} =\frac{3}{2}^{-} $  and $J^{P} =\frac{5}{2}^{+} $. Here we note that using the experimental data for the mass of $\frac{3}{2}^{-}$ and $\frac{5}{2}^{+}$  states in our sum rules we find the residues
$ \lambda_{\frac{3}{2}^{-}} =(0.98\pm0.05)\times 10^{-3}~\mathrm{GeV}^{6}$ and $ \lambda_{\frac{5}{2}^{+}} =(1.02\pm0.23)\times 10^{-3}~\mathrm{GeV}^{6}$, which are very close to the related values in the table \ref{tab:Values} and we do not see considerable differences. 
 Our results on the masses of the opposite-parity states, i.e.,  $\frac{3}{2}^{+}$  and  $\frac{5}{2}^{-}$  as well as our predictions on the residues may be verified via different  approaches.

\section{Summary and Outlook}

We performed  QCD sum rules  analyses to compute the mass and residue of the  hidden-charm  pentaquark states with $J=\frac{3}{2} $ and $J =\frac{5}{2} $ and both the positive and negative parities. We adopted interpolating  currents in an  anti-charmed meson-charmed baryon molecular form of $ \bar{D}^*\Sigma_{c} $ for states having $J =\frac{3}{2} $ and a mixed anti-charmed meson-charmed baryon molecular current of $[\bar{D}\Sigma_c^*]$ and $[\bar{D}^*\Lambda_c]$ for the states with $J=\frac{5}{2} $. By fixing the auxiliary parameters entered the calculations we obtained the values of the masses and residues for all the considered states.
Our predictions on the mass of the $J^{P} =\frac{3}{2}^{-} $ and $J^{P} =\frac{5}{2}^{+} $ states are consistent with the experimental data of the LHCb collaboration for the  masses of $P_c^+(4380)$ and $P_c^+(4450)$ states, respectively.  Our results are also consistent with the predictions of the theoretical works~\cite{Wang:2015epa,Chen:2015moa} on the masses. As we previously said Ref. \cite{Chen:2015moa} uses the same picture and method  with the present work but has prediction only for the masses of the $J^{P} =\frac{3}{2}^{-} $ and $J^{P} =\frac{5}{2}^{+} $  states. However,  Ref. \cite{Wang:2015epa} applies a different quark organization to predict also  the masses of the $J^{P} =\frac{3}{2}^{-} $ and $J^{P} =\frac{5}{2}^{+} $  states.
 
 By the adopted currents in the present study, we also derived the values of the residues for the considered states with both parities.  We got comparable results on the residues of $J^{P} =\frac{3}{2}^{-} $ and $J^{P} =\frac{5}{2}^{+} $  states with those of  Ref. \cite{Wang:2015epa}  within the errors.  The residues can be used as  the main inputs in the analyses of the electromagnetic properties and strong decays of the pentaquark states 
$P_c^+(4380)$ and $P_c^+(4450)$. Such analyses are needed and would  be very important in determination of the internal structures, geometric shapes, charge distribution, multipole moments of these states  and strong interactions inside them. In our future works we aim to analyze the strong, electromagnetic and weak decay channels of the pentaquark states considered in the present study  to calculate the corresponding strong coupling constants as well as the widths of these states. Comparison of the theoretical results on many parameters of the pentaquarks with the present and future experimental data  would help us better understand their quark organizations  and will provide us with useful knowledge on the quantum chromodynamics of the exotic baryons.


\section*{ACKNOWLEDGEMENTS}

This work was supported by T\"{U}B\.{I}TAK under the Grant no: 115F183.


\appendix*

\section{ The two-point spectral densities}

\renewcommand{\theequation}{\Alph{section}.\arabic{equation}}
\label{sec:App}
In this appendix we present the results for the two-point spectral
densities obtained from QCD sum rules calculations. As examples, we only present those spectral densities corresponding to the structures $g_{\mu\nu}$  and  $  \frac{g_{\mu\rho}g_{\nu\sigma}+g_{\mu\sigma}g_{\nu\rho}}{2} $ for  the states with $ J= \frac{3}{2}$ and $ J= \frac{5}{2}$, respectively. They are obtained  as

\begin{equation}
\rho_{i} ^{\mathrm{QCD}}(s)=\rho_{i} ^{\mathrm{pert.}}(s)+\sum_{k=3}^{6}\rho_{i,k}(s),
\label{eq:A1}
\end{equation}%
with $ i $ being $ \frac{3}{2}, \frac{5}{2} $.  In Eq.\ (\ref{eq:A1}) by $\rho_{i,k}(s)$    we
denote the nonperturbative contributions to spectral densities $\rho_{i} ^{\mathrm{QCD}}(s)$. The
explicit expressions for $\rho_{i} ^{\mathrm{pert}}(s)$ and $\rho_{i,k}(s)$ are
obtained  in terms of  the integrals of the Feynman parameters $x$ and $ y $ as:
\begin{widetext}
\begin{eqnarray} 
\rho_{\frac{3}{2}}^{\mathrm{pert}}(s)&=&\frac{m_c}{5\times 2^{15}\pi ^{8}}\int\limits_{0}^{1} dx\int\limits_{0}^{1-x}dy \frac{\left(6hsxy-m_c^2r \right)\left(hsxy-m_c^2r \right) ^{4} }{h^3 t^8} \Theta\left[L \right] , 
 \notag \\
\rho_{\frac{3}{2},\mathrm{3}}(s)&=&\frac{m_c^2}{2^{9} \pi ^6}\langle
\bar{d}d\rangle\int\limits_{0}^{1} dx\int\limits_{0}^{1-x}dy \frac{\left( hsxy-m_c^2t(x+y)\right) ^{3}}{h^2t^5}\Theta\left[L \right],  
   \notag \\
\rho_{\frac{3}{2},\mathrm{4}}(s)&=&\langle\frac{\alpha_{s}}{\pi}G^{2} \rangle \frac{1}{3^{2} \times 2^{15} \pi ^6}\int\limits_{0}^{1} dx\int\limits_{0}^{1-x}dy \frac{\left[hsxy-m_c^2t(x+y) \right] }{h^3 t^7}\left\lbrace 12 h m_c s xy^3(h^2 s x^3+m_c^2 t^2 y)\right. 
 \notag \\
&-& \left. 6m_c y \left( m_c^2 t (x+y)-hsxy\right)\left[2h^2 sx^3y+m_c^2t^2y^2+hsx \left(34x^4+2y(y-1)^2(16y-9)
\right. \right.\right.
 \notag \\
&+& \left.\left.\left.
x^3(105y-88)+x^2(72-209y+137y^2)+2x(50y^3-102y^2+61y-9) \right)   \right]
\right.
 \notag \\
&+& \left.
m_c\left(hsxy-m_c^2t(x+y) \right) ^2 \left[6h^2y^2+\left( 68x^4+3y(y-1)^2(17y-12)
\right.\right.\right.
 \notag \\
&+& \left.\left.\left.
x^3(197y-176)+8x^2(18-49y+31y^2)+3x(58y^3-123y^2+77y-12)\right)  \right]   
 \right\rbrace  
\Theta\left[L \right],  \notag \\
\rho_{\frac{3}{2},\mathrm{5}}(s)&=&\frac{3m_c^2}{2^{10} \pi^6}m_0^2\langle
\bar{d}d\rangle\int\limits_{0}^{1} dx\int\limits_{0}^{1-x}dy\frac{\left( hsxy-m_c^2t(x+y)\right)^{2} }{ht^4}\Theta\left[L \right],
\notag \\
\rho_{\frac{3}{2},\mathrm{6}}(s)&=& \frac{m_c}{3^3\times 2^{8}\pi^6}\left(2g_s^{2} \langle\bar{u}u\rangle^{2}+g_s^{2} \langle\bar{d}d\rangle^{2} \right) \int\limits_{0}^{1} dx\int\limits_{0}^{1-x}dy \frac{x\left(m_c^2r- 3hsxy\right) \left(m_c^2r- hsxy\right)}{t^5} \Theta\left[L \right]
\notag \\
&+&\frac{m_c}{ 2^{4}\pi^4}\langle\bar{u}u\rangle^{2}\int\limits_{0}^{1} dx\int\limits_{0}^{1-x}dy \frac{x\left(m_c^2r- 3hsxy\right) \left(m_c^2r- hsxy\right)}{t^5} \Theta\left[L \right],\nonumber\\
\rho_{\frac{5}{2}} ^{\mathrm{pert}}(s)&=&\frac{m_c\left(5 \cos^2\theta-4 \cos \theta \sin\theta+12 \sin^2\theta \right) }{2^{17}\times 3 \times 5^{2}\pi ^{8}}\int\limits_{0}^{1} dx\int\limits_{0}^{1-x}dy  \frac{ x \left(5 x^2+x (y+5z)+ 5 z y\right)}{h^3 t^9}
\nonumber\\
&\times&
 \left(s x yh-m_c^2 r\right)^4
\left(m_c^2 r-
6 s x yh\right)\Theta\left[L \right] ,  \notag \\
\rho _{\frac{5}{2},\mathrm{3}}(s)&=&-\frac{m_c^2\left(\cos^{2}\theta
   (\langle
\overline{d}d\rangle+4 \langle
\overline{u}u\rangle)+4 \cos \theta \sin \theta (\langle
\overline{d}d\rangle-2
   \langle
\overline{u}u\rangle)\right)}{2^{11}\times 3^{2} \times \pi ^6}\int\limits_{0}^{1} dx\int\limits_{0}^{1-x}dy\frac{ \left(3 x^2+x (y+3z)+3yz\right) }{h^2 t^6} \nonumber\\
&\times&\left(m_c^2 r-s x y h\right)^3\Theta\left[L \right],
   \notag \\
\rho _{\frac{5}{2},\mathrm{4}}(s)&=&-\langle\frac{\alpha_{s}}{\pi}G^{2} \rangle \frac{m_c}{2^{17}\times 3^{3}\times 5 \pi ^6}\int\limits_{0}^{1} dx\int\limits_{0}^{1-x}dy\frac{x (m_c^2 r-s x y h) }{h^3t^8}
\Bigg\{4 \cos \theta \sin \theta\Big(4 s^2 x^2 y^2 h^2 \Big(20 x^6 + 100 z^3 y^3
\nonumber\\
&+&
 4 x^5 (31 y+10z) + 5 x z^2 y^2 (56z + 27 y)
     +40 x^3 y (13 - 33 y + 20 y^2) + x^4 (20 - 504 y + 505 y^2) +5 x^2 y
\nonumber\\ &\times&
      (219 y - 337 y^2 + 154 y^3-36)\Big) +
  m_c^4 t^2 \Big (20 x^8 +
     10 z^2 y^5 (22z + 3 y) + x^7 (40z + 314 y) +
     x^6 (20 - 1004 y )
\nonumber\\ &+&
    1639 y^2 +
     2 x^5 y (475 - 2192 y + 1956 y^2) +
     3 x y^4 (1095 y - 1232 y^2 + 457 y^3-320 )
    + x^2 y^3 ( 6525 y-8537y^{2}
    \nonumber\\ &+&
     3572 y^3-1560) +
     x^3 y^2 (-1120 + 6865 y - 11362 y^2 + 5623 y^3) +
     x^4 y (-300 + 3865 y - 9221 y^2 + 5779 y^3)\Big)
\nonumber\\ &-&
  m_c^2 s x y \Big(100 x^{10} + 10 z^4 y^5 (62z + 9 y) +
     10 x^9 (40y + 93 y) +
     x z^3 y^4 (2880 - 7505 y + 4733 y^2)
\nonumber\\ &+&
     x^8 (600 - 6220 y + 6633 y^2) +
     x^2 z^2 y^3 ( 23645 y -4920- 34196 y^2 +  15489 y^3) +
     x^7 (  11700 y -400
    \nonumber\\ &-&
      29884 y^2 + 18857 y^3) +
     x^3 z^2 y^2 (-3680 + 29025 y - 56766 y^2 + 31998 y^3) +
     2 x^6 (50 - 5540 y
     \nonumber\\ &+&
      26777 y^2 - 39055 y^3 + 17777 y^4) +
     2 x^5 y (2645 - 23834 y + 63097 y^2 - 65643 y^3 + 23735 y^4) \nonumber\\ &+&
     x^4 y (-1020 + 21045 y - 98406 y^2 + 183623 y^3 - 151144 y^4 +
        45902 y^5))\Big)\nonumber
   \end{eqnarray}
\begin{eqnarray} 
       &+& 24 \sin^{2} \theta \Big(m_c^4 t^2 (20 x^8 + x^7 (83z-17) -
    15 z^2 y^4 ( 3 y^2-2)+
    x^3 y (120 - 750 y + 895 y^2 + 256 y^3 - 524 y^4) \nonumber\\&+& x^6 (170 - 398 y + 113 y^2) -
    x^5 (120 - 665 y + 623 y^2 + 36 y^3)+x^2 y^2 (180 - 630 y + 345 y^2 + 541 y^3 - 436 y^4)  \nonumber\\&+&
    x^4 (30 - 470 y + 1080 y^2 - 347 y^3 - 332 y^4) +
    x y^3 (120 - 290 y+20 y^2 + 353 y^3 - 203 y^4)\Big)  \nonumber\\&+&
 4 s^2 x^2 y^2 h^2 \Big(20 x^6 - 30 z^3 y^2 +
    4 x^5 (22 z-3) + 10 x z^2 y (6 - 11 y + y^2) +
    x^4 (170 - 318 y + 145 y^2)
\nonumber\\
&+&
    5 x^3 (-24 + 86 y - 89 y^2 + 27 y^3) +
    10 x^2 (3 - 26 y + 50 y^2 - 33 y^3 + 6 y^4)\Big)\nonumber\\&-&
 m_c^2 s x y \Big(100 x^{10} + 35 x^9 ( 19 z-1) -
    15 z^4 y^4 ( 8 y + 5 y^2-10) +
    6 x^8 (325- 690 y + 336 y^2)\nonumber\\
&-&
    2 x z^3 y^3 (300 - 740 y + 300 y^2 + 167 y^3) +
    2 x^7 (-1400 + 5225 y - 5704 y^2 + 1837 y^3) \nonumber\\&-&
    x^3 z^2 y ( 5580 y - 11835 y^2 + 6772 y^3 +
       319 y^4-600) -
    x^2 z^2 y^2 ( 4620 y - 6505 y^2 2152 y^3 +
       642 y^4\nonumber\\&-&900 ) +
    x^6 (2200 - 13760 y + 26198 y^2 - 19000 y^3 + 4353 y^4) +
    x^5 (10005 y - 31216 y^2+39533 y^3-900 \nonumber\\&-&   20672 y^4 +
       3250 y^5) +
    2 x^4 (75 - 1910 y + 10145 y^2 - 20991 y^3 + 19413 y^4 -
       7324 y^5 + 592 y^6)\Big)\Big)\nonumber\\&+& 
       5 \cos^2\theta\Big(4 s^2 x^2 y^2 h^2 \Big(52 x^6 +
     4 z^3 y^2 ( 5 z-13) + 4 x^5 ( 61 z-1) +
     x z^2 y (144-320 y + 107 y^2) +x^4\nonumber\\&\times&
     (412 - 864 y + 449 y^2) -
     4 x^3 (72 - 284 y + 333 y^2 - 121 y^3) +
     x^2 (72 - 660 y + 1419 y^2 - 1129 y^3 + 298 y^4)\Big) \nonumber\\&+&
  m_c^4 t^2 \Big(52 x^8 +
     x^7 ( 270 z+22) - 2 z^2 y^4 (  22 y + 29 y^2) +
     x^6 (412 - 1156 y + 599 y^2-36) \nonumber\\&+&
     2 x^5 ( 893 y - 1186 y^2 + 348 y^3-144) +
     x^2 y^2 (432 - 1824 y + 2133 y^2 - 409 y^3 - 332 y^4) \nonumber\\&+&
     x^3 y (288 - 2024 y + 3521 y^2 - 1658 y^3 - 133 y^4) -
     3 x y^3 (296 y - 235 y^2 -36 y^3 + 71 y^4-96) \nonumber\\&+&
     x^4 (72 - 1188 y + 3365 y^2 - 2677 y^3 + 359 y^4)) -
  m_c^2 s x y (260 x^{10} + 2 x^9 (-880 + 931 y) \nonumber\\&-&
     2 z^4 y^4 (  206 y + 19 y^2-180) +
     5 x^8 (960 - 2236 y + 1233 y^2) +
     x z^3 y^3 ( 4128 y -2941 y^2-1440 \nonumber\\&+&   145 y^3) +
     x^7 (  27420 y - 33356 y^2 + 12589 y^3-6800) +
     x^2 z^2 y^2 (2160 - 12072 y + 20341 y^2-12004y^3 \nonumber\\&+&
        1557 y^4) +
     x^3 z^2 y (1440 - 14128 y + 34209 y^2 - 27606 y^3 +
        5634 y^4) +
     2 x^6 (2650 - 17620 y \nonumber\\&+& 36793 y^2 - 30611 y^3 + 8779 y^4) +
     2 x^5 (  12535 y - 42226 y^2 + 60059 y^3 - 37935 y^4 +
        8647 y^5-1080) \nonumber\\&+&
     x^4 (360 - 9372 y + 52905 y^2 - 120438 y^3 + 129907 y^4 -
        65384 y^5 + 12022 y^6)\Big)\Big)\Bigg\} \Theta\left[L \right] ,  \notag \\\nonumber\
   \end{eqnarray}
\begin{eqnarray} 
\rho _{\frac{5}{2},\mathrm{5}}(s)&=&\frac{\left( \cos \theta-2\sin \theta\right) m_c^2 m_0^2\left( 6 \sin \theta \langle
\overline{d}d\rangle + \cos \theta (\langle
\overline{d}d\rangle +4 \langle
\overline{u}u\rangle)\right) }{2^{13} \pi^6}\int\limits_{0}^{1} dx\int\limits_{0}^{1-x}dy\frac{\left( s x y h-m_c^2 t (x+y)\right)^{2} }{h t^5}
\nonumber \\
&\times&\left( 2x^2 +x(3z+1)+2yz\right) 
\Theta\left[L \right],\nonumber\\
\rho _{\frac{5}{2},\mathrm{6}}(s)&=&\int\limits_{0}^{1} dx\int\limits_{0}^{1-x}dy\left\lbrace\left( 2g_s^{2} \langle\overline{u}u\rangle^{2}+g_s^{2} \langle\overline{d}d\rangle^{2}\right)  \frac{m_c\left(5 \cos^2 \theta-4 \cos \theta \sin \theta+12 \sin^2 \theta\right) }{2^{11}\times3^{4} \pi^6}\left(m_c^2t(x+y)-3shxy \right)\right.
\nonumber\\
      &\times& \left.
\left(m_c^2t(x+y)-shxy \right)\left( 2xyz+x^2(2x+3y-2)\right)
- \frac{m_c\left(\cos \theta-2 \sin \theta \right)}{3\times 2^{8}\pi^4 t^5}\left[ \langle\overline{u}u\rangle^{2}  \left(\cos \theta+6\sin \theta \right) +4\langle\overline{u}u\rangle \langle\overline{d}d\rangle \cos \theta\right]  \right.
\nonumber\\
      &\times& \left.\left(  m_c^2tx(x+y)-3shx^2y\right)\left(m_c^2t(x+y)-shxy \right)   \right\rbrace \Theta\left[L \right],
\end{eqnarray}

\end{widetext}
where $ \Theta\left[L \right] $ is the usual unit-step function and we have  used the  shorthand notations 
\begin{eqnarray}
z&=&y-1,
\nonumber\\
h&=&x+y-1,
\nonumber\\
t&=&x^2+(x+y) (y-1),
\nonumber\\
r&=&x^3+x^2 (2 y-1)+ y(y-1)(2x+y),
\nonumber\\
L&=&\frac{z}{t^2}\left[sxyh-m_c^2(x+y)t\right] .
\end{eqnarray}

\end{document}